\documentclass[sigconf,authorversion]{acmart}

\usepackage{color}
\usepackage{subcaption}
\usepackage{hyperref}
\usepackage{graphicx}
\usepackage[moderate,margins=normal,leading=normal]{savetrees}
\usepackage[]{algorithm2e}

\newcommand{\scomment}{\textcolor{black} } 
\newcommand{\paragraphb}[1]{\vspace{0.03in}\noindent{\bf #1} }

\acmYear{2021}\copyrightyear{2021}
\setcopyright{acmlicensed}
\acmConference[WiSec '21]{Conference on Security and Privacy in Wireless and Mobile Networks}{June 28--July 2, 2021}{Abu Dhabi, United Arab Emirates}
\acmBooktitle{Conference on Security and Privacy in Wireless and Mobile Networks (WiSec '21), June 28--July 2, 2021, Abu Dhabi, United Arab Emirates}
\acmPrice{15.00}
\acmDOI{10.1145/3448300.3467816}
\acmISBN{978-1-4503-8349-3/21/06}
\begin{document}


\title{VoIPLoc: Passive VoIP call provenance via acoustic side-channels}

\author{Shishir Nagaraja, Ryan Shah}
\email{{first.last}@strath.ac.uk}
\affiliation{%
  \institution{University of Strathclyde}
  \country{UK}
}

\ccsdesc[500]{Networks~Web protocol security}
\ccsdesc[500]{Security and privacy~Pseudonymity, anonymity and untraceability}
\ccsdesc[500]{Networks~Network security}

\keywords{VoIP security, call provenance, source identification, location privacy, acoustic fingerprint}




\begin{abstract}

We propose VoIPLoc, a novel location fingerprinting technique and
apply it to the VoIP call provenance problem. It exploits echo-location
information embedded within VoIP audio to support fine-grained
location inference. We found consistent statistical features induced
by the echo-reflection characteristics of the location into recorded
speech. These features are discernible within traces received at the
VoIP destination, enabling location inference. We evaluated VoIPLoc by
developing a dataset of audio traces received through VoIP channels
over the Tor network. We show that recording locations can be
fingerprinted and detected remotely with a low false-positive rate, even
when a majority of the audio samples are unlabelled. Finally, we note
that the technique is fully passive and thus undetectable, unlike
prior art. VoIPLoc is robust to the impact of environmental noise and
background sounds, as well as the impact of compressive codecs and
network jitter. The technique is also highly scalable and offers
several degrees of freedom terms of the fingerprintable space.
\end{abstract}

\keywords{VoIP security, call provenance, source identification, location privacy, acoustic fingerprint}

\maketitle



\section{Introduction}

Voice-over-IP (VoIP) protocols enable millions of individuals to
communicate inexpensively regardless of geographic location. The
rising popularity of VoIP over the years, especially during the
Covid-19 pandemic, has also lead to its abuse by criminal
gangs. Cyber-criminals exploit the relative anonymity of VoIP clients
as the meta-data regarding caller location can be easily spoofed in
VoIP infrastructure. This has lead to efforts to identify the origin
of VoIP calls via coarse-grained techniques, such as those that
leverage route-specific characteristics of call
audio~\cite{bala:ccs:2010}, among others. Such efforts will give a
rough idea of where a caller is located down to a specific
city. VoIPLoc enables fine-grained call location provenance by
confirming whether or not a call was made from a specific room in a
building, if that room has been previously fingerprinted.

For users of anonymous communications, VoIPLoc location fingerprinting
is an attack of concern. Political dissidents coordinating a protest,
journalists covering an event, or business persons making deals
require location anonymity to ensure operational security. They often
use VoIP applications over Tor to hide endpoint information. While Tor
hides the source origin on the IP infrastructure, and mitigates
coarse-grained attacks, side-channel attacks are a constant worry.

\begin{figure}
	\centering
	\includegraphics[width=1\linewidth]{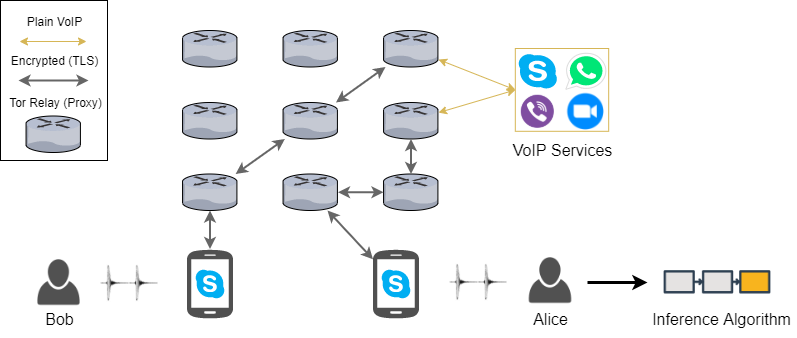}
	\caption{The VoIP-call Provenance Problem}
	\label{fig:workflow}
        \vspace{-2em}
\end{figure}

In this work, we demonstrate that VoIP channels inadvertantly
advertise unique location fingerprints in the acoustic channel, as a
fundamental property of recorded audio. This can be used to mount a sidechannel attack on Tor users using VoIP tools. As a concrete
contribution, we developed an attack that exploits this vulnerability
to establish call provenance, even over anonymous VoIP channels. The
attacker's goal is to establish call provenance -- {\em tracing the
original location of an audio source}, as illustrated in
Figure~\ref{fig:workflow}.

Our contributions are as follows. We describe and evaluate a
fully-passive location fingerprinting technique (VoIPLoc) for anoynmous VoIP
channels. We then demonstrate fine-grained tracking in call provenance. Finally,
we demonstrate our technique's properties of uniqueness, time-invariance, and
robustness to variance in network characteristics.

\section{Threat model}
\label{sec:threatmodel}
The threat model is that the adversary has access to the audio stream,
such as recorded speech, at a communication endpoint. For
example, an intelligence agency analysing streaming audio from a
dissident website, or tracing a dissident activist's location
during a phone interview. From a communications perspective, the
adversary is an insider engaged in a VoIP conversation with
the victim. Such a threat model is reasonable, under the assumption that
motivated adversaries would not restrict themselves to launching
external attacks.


One of the criticisms of the insider threat model is that it is
unrealistic. {\em Why would a victim hold a VoIP conversation with an
attacker?} A common counter-argument to this position is the following
stance: any motivated adversary will be an insider. As in most
real-world situations, trust isn't binary. A salesperson for a firm
dealing in radioactive materials, or surveillance equipment, might be
contacted by attackers. Posing as prospective clients, they can gain
information about the salesperson's clients after compromising their
location privacy. Snowden's revelations famously revealed the tracking
of sales personnel by tracking the victim's mobile
phone~\cite{guardiansnowden2013}. As opposed to the macro-level
location information provided by cell-tower localization, we report
attacks that can carry out fine-grained indoor-location
identification. Specifically, down to a specific room or corridor the
victim made a voice call in.

VoIPloc can be also used to identify the location of attackers
involved in running spam campaigns, remotely carrying out Covid-19 quarantine
check-ins, or other types of cybercrime and fraud. These types of attacks
engage victims via audio calls. Thus both legitimate and
illegitimate actors should be concerned about their location
privacy. The undetectability of fully passive location fingerprinting
attacks poses a significant threat to online privacy of VoIP
participants.

We have specifically chosen to study passive adversaries to
demonstrate the power of the attacks. Even a passive attacker can
confirm the source-location of an audio call. The passive attacker
model studies the lower bound of attacker success, given highly
restrictive circumstances for location identification. Speech codecs
apply bandpass filters to compress audio, which excludes techniques
that use inaudible frequency bands such as
ultrasound~\cite{cheng2018sonarsnoop} or LF/VLF band audio (eg. the buzz of
mains frequency). This forces active attacks to use audible frequencies,
lacking in stealth and thus showcasing the importance of passive attacks.


\vspace{-0.75em}
\subsection{Requirements}
Following the threat model, we now establish our requirements for
practical VoIP call provenance techniques:
\vspace{-0.8em}
\begin{enumerate}
  \item { {\em Stealth} -- Provenance should be established via
   passive methods, in order to prevent detection by call
   participants. With an
   obvious stealth advantage, passive attacks are less observable than active
   attacks. Further, they are more robust to the compressive effects of codecs,
   which may remove ultrasound and infrasound components due to the deployment
   of aggressive band-pass filters.}

  \item { {\em Fine-grained tracking} -- Provenance should be
   established down to the specific room used to make a call, thus
   affording the caller the smallest anonymity set.}

  \item{ {\em Uniqueness} -- A provenance fingerprint must be
   distinct.}

  \item{ {\em Time invariance} -- Provenance fingerprints should not
   rely on leveraging background sounds for fingerprinting. For example, relying
   on proximity to an external noise source that might be unavailabl can
   result in an unreliable fingerprint.}

  \item{ {\em Robustness} -- Provenance fingerprints should be
   robust to the presence of background noise, such as HVAC and
   fans. The source (i.e victim) should not be required to carry
   specialised hardware in order to reliably establish
   provenance.}
\end{enumerate}

\section{Attack Technique}

\begin{figure}[!htbp]
\begin{center}
 \includegraphics[width=0.75\linewidth]{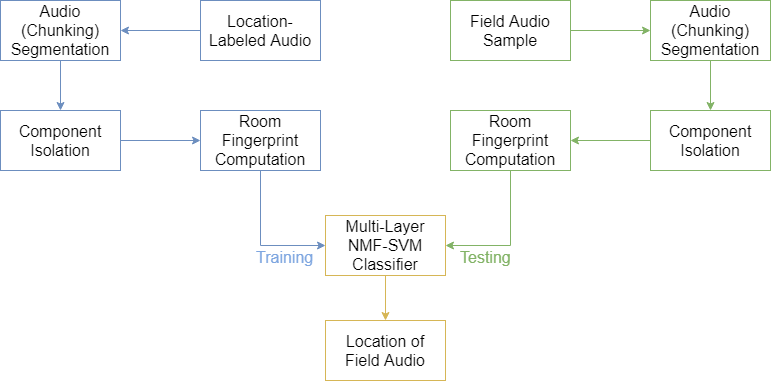}
\end{center}
 \vspace{-0.3em}
\caption{Attack workflow}
\label{fig:algflow}
\vspace{-2em}
\end{figure}

The attack workflow consisting of decomposition and classification is shown in
Figure~\ref{fig:algflow}. Location fingerprints are computed by
extracting acoustic-reflection characteristics of the location, whilst
filtering out all background sounds. A multi-layer (deep) classifier
then maps the fingerprint to a location by comparing it to a database
of previously pre-labeled audio samples.


%

\vspace{-1em}
\subsection{Preprocessing step -- Audio segmentation}
The initial step in the attack is the collection of audio traces by
the attacker, as an insider on a VoIP call. Thereafter, the attacker
proceeds to segment audio traces into chunks, such that each
chunk corresponds to a single {\em utterance} -- the smallest unit of
human speech with a brief silence at either end. We followed standard
practice of Ishizuka et al.~\cite{ishizuka:2010:sc} towards
segmentation, by passing the traces through a voice activity detector
and a silence detector~\cite{ramirez:2007:ieeetaslp}. Successive
bursts of spectral flatness, with a burst of spectral energy in between,
is used to differentiate between voice activity and silence.



\vspace{-1em}
\subsection{Multi-layer decomposition -- Isolating location-specific signal components}
\label{sec:latereverb}

The next step is to isolate the relevant signal component that is a
function of the location from each chunk.

\begin{figure}[!htbp]
\begin{center}
 \includegraphics[width=0.8\linewidth]{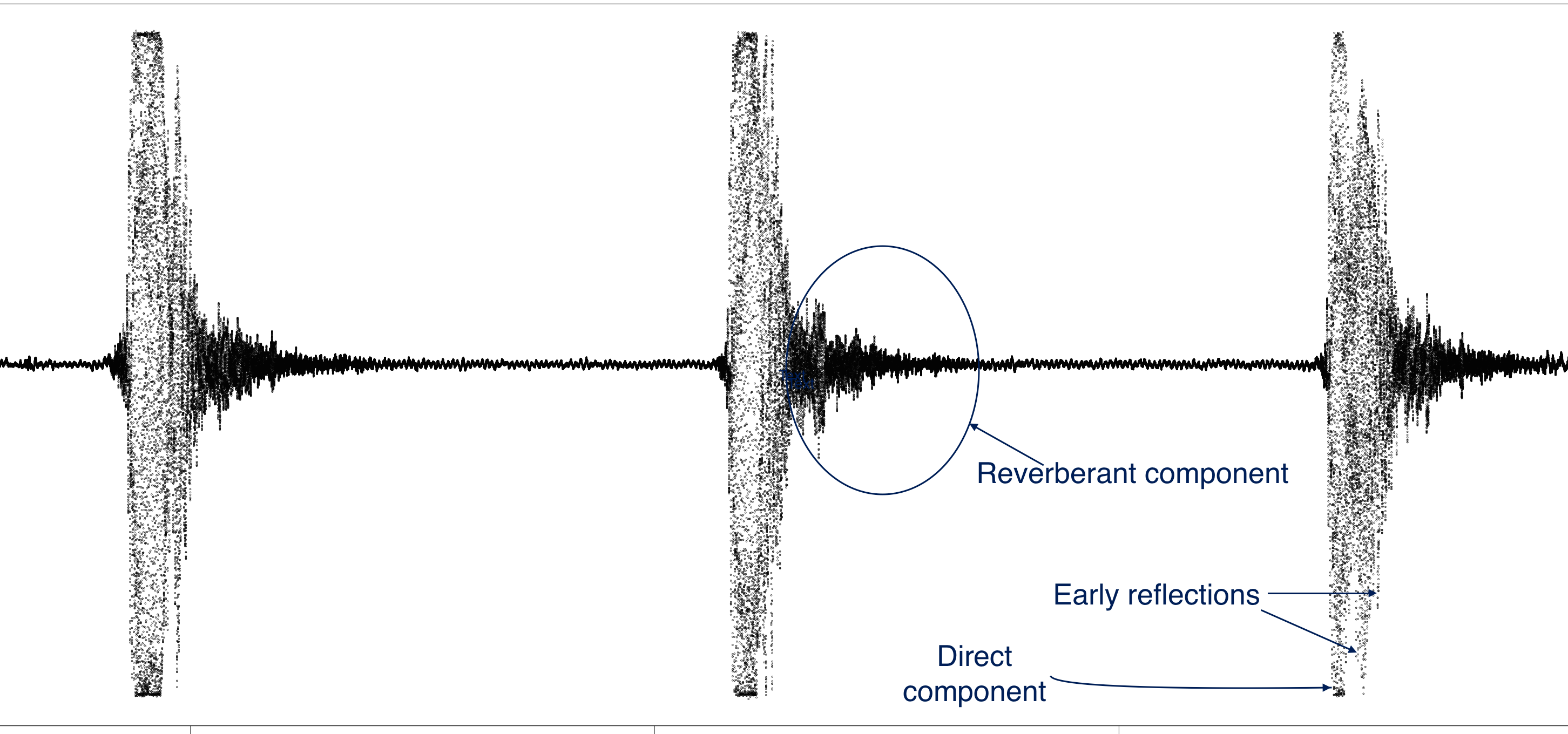}
\end{center}
\caption{Components of a sound sample}
\label{fig:rec}
\end{figure}

Each audio chunk is made from three signal components (see
Figure~\ref{fig:rec}). First, {\em direct sound} is the sound
transmitted in a direct path, from the speaker to the microphone with
no reflections. Second, {\em reflections} follow direct
sound. These are distinct reflected sounds that arrive at the VoIP
sender's microphone along a predictable path. Third, the
{\em reverberant component} is composed of higher order reflections, which
are a combined function of all the room surfaces. To fingerprint a
location, the reverberant component is the most relevant. This is because it is a
stable function of acoustic information, diffused throughout the
location. Isolating the reverberant component is non-trivial due to the
time-overlapping nature of the three components.

As the speaker utters a sound, the direct-sound components
overlaps with the early-reflections and reverberation components from
previously spoken syllables. This is because the reflections from a
previously spoken syllable are still above the noise floor when the
current syllable is uttered. The challenge of isolating reverberant
component is therefore to decompose a given speech segment back into
the reverberant component, early reflections, and direct-sound.
Since we are only interested in the reverberant component (which is a
function of the room rather than the speaker), it is critical that
any direct-sound components and early reflections are removed from
the reverberant component. This is a technical challenge that a
fine-grained fingerprinting technique using human voice must solve.

\begin{figure}[!htbp]
\begin{center}
 \includegraphics[width=1\linewidth]{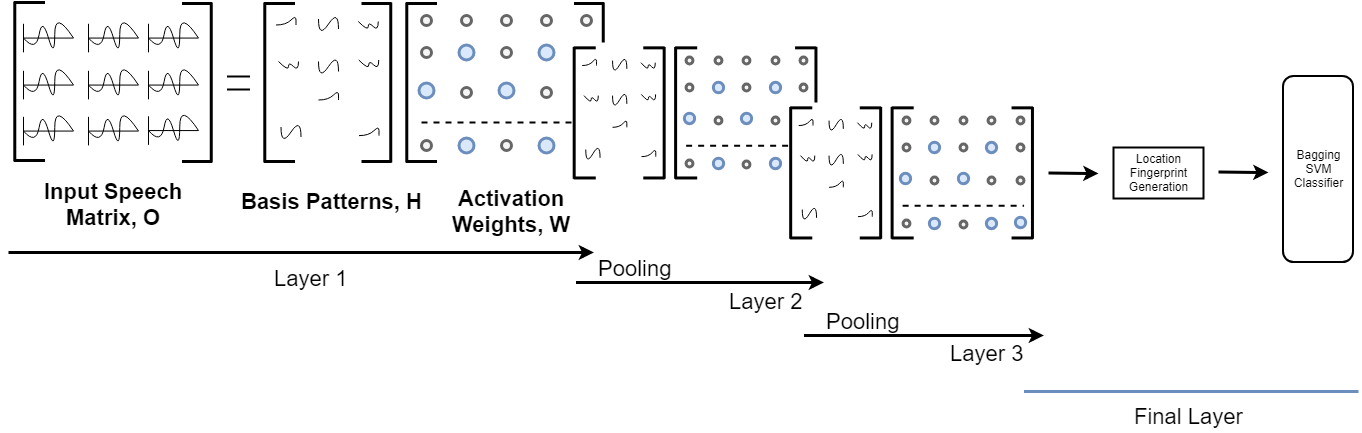}
\end{center}
 \vspace{-1em}
 \caption{Multi-layer Inference Algorithm}
  \vspace{-2em}
\label{fig:alg}
\end{figure}

\begin{figure}[!htbp]
\vspace{-0.5em}
\begin{center}
 \includegraphics[width=0.80\linewidth]{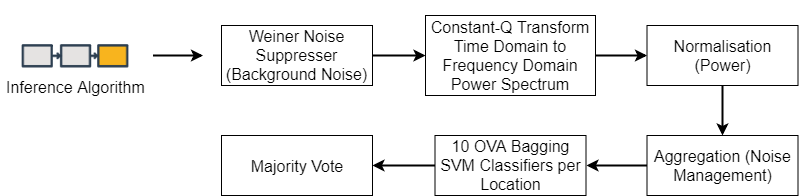}
\end{center}
 \vspace{-0.3em}
\caption{Fingerprint computation after Inference}
\label{fig:fingerprint}
 \vspace{-1em}
\end{figure}

To isolate overlapping components and suppress background noise
sources, VoIPLoc uses a compressive decomposition technique
(Figure~\ref{fig:alg}), which partitions recorded audio into direct
sound, reflections, and the reverberant component. Partitioning
leverages the observation that the shape and form of reflections are,
by definition, very similar to direct sound. As such, early
reflections can be constructed using the same direct sound and the
addition of appropriate location-specific transformation. Partitioning
has three stages: (a){\em $K$ layers of nested NMF decomposition with
 pooling}; (b) {\em Fingerprint generation function}; and (c) the
{\em non-linear classifier}.

\subsubsection{Decomposition}
The decomposition approach in this section was originally developed by
Nagaraja et al.~\cite{nagaraja:wisec:19} for binary analysis. We
extend this to the non-binary case and significantly augment it.

In preparation for decomposition, the attacker loads audio traces into
the audio matrix $O$, where each row correponds to one chunk. Each
chunk is split into $t$ time intervals. Within each time interval the
attacker computes the average signal amplitude, thus yeilding a vector
of signal amplitudes (in dB).

First, input traces are expressed as a linear combination of direct
sound and location-specific transformations ($H$), and mixing weights
($W$) as: \(O_{ij} = \sum_{k=1}^K H_{ik} W_{kj}\). The key intuition
here is to represent speech segments using the sparsest and fewest
number of speech sources. Maximising sparsity is the basis for
compressing a signal containing the direct sound and their reflections
back into a handful of direct-sound patterns and the intensities of
reflection. Within each layer, this is governed by the {\em
  optimisation} function which is formally stated as: \( \min ||O^1 -
H^1W^1||\) such that $W^1_{ij} \geq 0, H^1_{ij} \geq 0$, i.e. $W^1$
and $H^1$ are non-negative, hence the name Non-Negative Matrix
factorisation, where $||\odot||$ is the Frobenius norm.  {\em
  Optimisation} is carried out by starting out with randomly
initialised positive-valued matrices $W^1$ and $H^1$, and updating
them iteratively using (standard practice) multiplicative update
rules~\cite{nips00:leeseung}: \( H_{ir} = \frac{ H_{ir} \sum_r
  \frac{O_{ij}}{(HW)_{ij}}H_{rj} }{\sum_l H_{jr}}\) and \( W_{rj} =
W_{rj} \sum_i H_{ir} \frac{O_{ij}}{(HW)_{ij}}\). By deploying multiple
layers of decomposition we can further promote sparsity. This motivates
the use of Deep Neural Networks (DNN)~\cite{alg:deep}. In each layer,
sound components are partitioned into sub-components,
sub-sub-components, and so on.

Second, we use pooling to build robustness to the position and
movement of the speaker. Thus each decomposition layer is followed by
max-pooling, a moving-window function which takes the rows of the
weight matrix $W^k$ as input, and replaces a subset of the row by the
maximum value of the subset. This approach was first suggested by
Boureau et al.~\cite{alg:pooling}. The pooling function for row $i$ of
weight matrix $W$ in layer $k$ as \(F(W^k_{i*}) = \max {W^k_{ip} | j-c
  \leq p \leq j+c, 0 \leq j \leq N} \), where $N$ is the number of
columns of the input audio-traffic matrix $O^1$ and $c$ is a
constant. Accordingly, $F(W^k) = (F(W^k_{1*},\dots,W^k_{M*} )$ is the
pooling function for matrix $W^k$. In our evaluation, we used a value
of $c=20$, which corresponds to a moving window of 20Hz.
Upon convergence, the columns of $H^1$, with normalised weight $\sum
W^X_{i*}/\sum W^X$ greater than $0.9$, contain direct-sound signals,
while the rest correspond to the reverberant component, where $W^X =
H^2 \dots H^K W^K$.

The final step is to isolate the reverberant component. We regenerate
the sound signal using all but the patterns corresponding to direct
sound in $H^1$. The columns of $H$ corresponding to direct sound are
set to zero and the signal $O_r$ is regenerated $O'=H^1W^X$. Non-zero
rows of $O'$ contain the reverberant component.

An important system parameter is the choice of number of columns of
$H$, i.e. the number of basis patterns. This should roughly be set to
constant times the number of possible sound sources that are
simultaneously active in a location. For instance, if victim is a
single VoIP caller sitting alone in a sound-proof room, then the
number of sources should be at least 5 (one for the speaker, and a few
for the reverberant signal and noise), there is no upper limit. We set
the number heuristically at $100$ in all our experiments.

\vspace{-1em}
\subsection{Fingerprint computation}


Fingerprint computation (Figure~\ref{fig:fingerprint}) consists of several steps: noise suppression, signal aggregation and classification.

\subsubsection{Noise suppression}
The reverberant component computed thus far contains transformation
noise (from the reflections) and environmental noise, both of which
must be removed. We used a Wiener noise
suppressor~\cite{plapous:2006:ieeetaslp}, to remove the influence of
background noise on the location fingerprint. This approach uses harmonic
regeneration noise reduction (HRNR), to refine the signal-to-noise
ratio before applying spectral gain to preserve speech harmonics.

\subsubsection{Reflection measurement}
The key idea underlying location fingerprint computation is to
compute the signal power in each frequency band, within the reverberant
component normalised by the corresponding signal power within the
direct sound components i.e we measure signal
attenuation as a function of room geometry.

\if false
To design a fingerprint function, one design consideration is the
choice of frequencies to consider. It is worth noting that the
correlation of signal power for different locations within a given
room is inversely related to the frequency. High frequencies are only
correlated at close proximity, while low frequencies (above Schroeder
frequency) can be correlated anywhere within the given
location~\cite{ramirez:2007:ieeetaslp}. Thus, signal power at lower
frequencies (20Hz --- 2KHz) is ideal for location fingerprinting.
\fi

The standard tool for frequency analysis is the Fourier transform. For
digital signals, the textbook approach to ascertain signal power by
frequency band is to apply the Discrete Fourier Transform; often using
the Fast Fourier Transform (FFT) algorithm. However, FFT is unsuitable
for our purpose. The average minimum (fundamental) frequency for human
speech varies from 80 to 260 Hertz; 85 to 180 Hertz for Basal and
Tenor voices and from 165 to 255 Hertz for Contral to Soprano
voices. Therefore, using FFT the frequency resolution would be
insufficient. FFT with 512 temporal samples recorded at a sampling
rate of 44.1 KHz, has resolution of 86.1 Hz between two FFT
samples. This is not sufficient for low frequencies found in human
voice. For instance, the distance between two adjacent vocal tones
could be as low as 8 Hz to 16 Hz. The frequency resolution can be
improved by using a higher number of FFT samples. For instance, with
8192 temporal samples, the resolution will be improved to 5.4Hz for a
sampling rate of 44.1 KHz. However, this alone is inadequate since the
signal at higher frequencies will have better resolution than those at
lower frequencies.

To provide constant frequency-to-resolution ratio for each frequency
band, we use the Constant-Q transform~\cite{brown:1991:JASA}. This is
similar to the Discrete Fourier Transform but with a crucial
difference -- it permits the use of a variable window width to achieve
constant resolution, enabling effective coverage across the
spectrum. Constant resolution is achieved via a logarithmic frequency
scale. The CQT transform of the reverberant component is computed,
after due isolation using the technique described in
Section~\ref{sec:latereverb}. The output of the transform is the {\em fingerprint
vector} which contains the signal power in each frequency band.

\vspace{-0.5em}
\subsubsection{Normalisation}
The fingerprint vector is normalised to remove biases arising from
variability in the input-signal amplitude; some speakers speak louder
while some speak softly, amplitude variance can also arise from
speaker movement. The fingerprint vector computed thus far is
normalised by the signal amplitude of the direct sound component in
the corresponding band via element-wise division of the CQT transform
of the reverberant signal ($R$) by the CQT transform of the
direct-sound signal ($D$). For each speech segment $i$, matrix
$R_{i*}$ is the CQT transform vector of the reverberant component, and
matrix $D_{i*}$ stores the CQT transform vector of the direct-sound
component.
We then aggregate the normalised vectors from each speech segment, to maximise the range of frequencies that can be used in the fingerprint. Thus we merge multiple CQT vectors computed over respective reverberation components, as follows. Vector $p$ contains the normalised aggregated fingerprint. For each segment $i$ and frequency $j$, we compute \( P_{ij} = \frac{R_{ij}}{D_{ij}} | D_{ij} > 0 \) and \( P_{ij} = 0 \forall D_i \leq 0 \). The signal power is then added up \(p_j = \sum_i^n P_{ij} \).
%
%

\vspace{-1em}
\subsection{Final Layer -- Non-linear separation}
\label{sec:classification}
The final layer is a supervised classification layer that maps the
outputs of the fingperprint generator onto a non-linear
multi-dimensional space. The fingerprint vector is input into an
ensemble of weak {\em bagging} classifiers. Their role is to map the
input to a location label. The classifier is trained positively with
location traces and negatively against traces from other locations.

\paragraphb{Pre-filtering for contamination-resistance:} It is
important to ensure that negative training sets are minimally
contaminated with positive samples to avoid overlapping label
definitions. To achieve this, we applied the (parameter free) Xmeans
algorithm to cluster vectors in the negative training set on a per-set
basis. Any cluster with labelled positive fingerprints is discarded
from the negative training set.


\paragraphb{Classifier choice:} We chose a non-linear kernel function
within an SVM classifier. The kernel function uses exponentiation of
the Euclidean distance to ensure linearity in the locality of a
fingerprint vector: \(K(x_i, x_j) = e^{(-\gamma ||x_i-x_j||^2)}\),
where $\gamma$ is the width of the Gaussian function. Our classifier
choice is driven by the following reasons: first, the dataset is
{\em small} as only a few samples per room-location are likely to be
available for training. Second, the dataset is {\em sparse} due to the
high dimensionality of fingerprint vectors. Thus an classifier like
SVM will have no problem identifiing separating hyperplanes that
maximises the {\em margin of separation} between data vectors $p$ with
a low overlap across class boundaries. Third, the classifier must be
robust to the presence of unlabelled data during testing. For
instance, in the context of deanonymising a VoIP caller using an
anonymous communication channel, a call could be made from a number of
locations that are not in the fingerprint database. This requirement
can be addressed within an SVM framework by using a bagging approach
-- instead of using just one classifier, we train ten classifiers per
location, in One-vs-All (OVA) mode where each of the ten classifiers
for the $i^{th}$ room are trained using a tenth of the samples from
the training set for the $i^{th}$ room with a positive label, and a
randomly chosen tenth of all the remaining samples (of the training
set including unlabelled samples) with a negative room label. In other
words, random resamples (of a 10th) of the unlabelled data are drawn
and the classifiers are trained to discriminate the positive room
sample from each resample. Resampling unlabelled locations induces
variability in classifier performance which the aggregation procedure,
used to combine the outputs of individual classifiers, can then
exploit. {\bf Aggregation:} The results of the classifiers for
$i^{th}$ room are aggregated with a majority vote. Overall, for $n$
rooms we train a total of $n \times 10$ classifiers, which is still
$O(n)$ classifiers. This method of combining classifier output is
based on the technique first introduced by Mordelet and
Vert~\cite{mordelet:2014:pr}.

\vspace{-1em}
\section{Evaluation}
\label{sec:eval}
In this section, we evaluate the attack technique using VoIP conversations in
a diverse set of locations, codecs, network jitter, and speech
characteristics using a corpus of recordings.

\vspace{-1em}
\subsection{Real-world dataset}
\label{sec:rdata}
Our first dataset consists of audio recordings of VoIP sessions
conducted over the Tor network, from 79 rooms of identical geometry of
a university computer science department. Occupants customise these
rooms using furnishings such as desks, bookshelves, monitors, and
other objects that affect room acoustics but are otherwise
identical. The impact of these customizations on the reverberant
component of recorded audio forms the basis for location
identification. Our goal is to understand the extent to which our
VoIPLoc exploits differences in acoustic absorption/reflection
characteristics whilst tolerating acoustic and network jitter. The
rooms have typical (acoustic) noise sources which could be continuous
such as air conditioning systems, heater fans, and fridges, or
intermittent noise from road traffic or human subjects in the
vicinity. The dataset was generated in 2016 and used the public Tor
network for experiments.




A VoIP session (see Figure~\ref{fig:workflow}) is set up over the Tor
network between the sender (UK) at the given location and a recipient on
the other end (Davis, CA, USA). At the receiver the resulting audio stream is
recorded. We recruited sixteen volunteers to conduct VoIP sessions in
each of the 79 rooms and recorded the audio at the recipient end. The
(sender) volunteers were selected for diversity in voice pitch ({\em 8 male and
 8 female}). For each room, we seated each of the volunteers at nine
different positions located at the intersections of a $3$x$3$ grid
(rectangular) control for position-specific bias.

Volunteers were instructed to remain in a neutral tone and hold a
conversation, moving naturally, whilst reading out from a script from
the NXT Switchboard Corpus~\cite{nxtcorpus} consisting of telephone
conversations between speakers of American English. It is one of the
longest-standing corpora of fully spontaneous speech. We used the
MS-State transcript of the corpus, and all volunteers read the same
transcript for consistency. The corpus has transcripts that support
conversations of different lengths for diversity.

\paragraphb{Training and testing sets:} Our dataset contains a
significant number of audio traces (288) per location, most of which
are utilised for testing and a small fraction are used for training (a
credible attack cannot depend on more than a couple of audio
samples). We partitioned the dataset into $k$ different
non-intersecting sets (by random allocation with uniform probability)
on a per-location basis. For each location, one set is used for
training and $k-1$ are used for testing, this is repeated $k=50$ times
so that every subset is used for testing (standard $k$ foldover
cross-validation).

\vspace{-1em}
\subsection{Codecs}
A wide variety of VoIP clients are in popular use thus we are
interested in the impact of speech codecs on
fingerprinting. Codecs apply a range of techniques such as compression
and variable sampling rates to efficiently encode as much of the
speech information as possible under assumed steady state network
conditions. The resulting compression presents a significant challenge
for remote fingerprinting due to the potential loss of relevant signal
information. Indeed many codecs apply a variable cutoff {\em high-pass filter} to remove ambient sounds -- low frequency background sounds and breathing noise.

\paragraphb{SILK codec (Skype):}
The widely used Skype VoIP services uses SILK codec~\cite{silk} as
well as other proprietary voice codecs for encoding high frequencies
in the range of 16KHz. The key parameter for our purposes is the target
bitrate. SILK's signal bandwidth (frequency range) varies in time
depending on network conditions. When throughput is low, lower
bitrates are used, and the codec enters a Narrowband mode wherein the
sampling rate is set to 8KHz and the signal bandwidth covers
300-3400Hz. In this mode, higher frequencies are not transmitted,
potentially affecting the performance of our fingerprinting
techniques. The range of frequencies skipped in this manner depends on
the network throughput. Internally, SILK supports 8, 12, 16, and 25KHz
resulting in bitrates from 6 to 40 Kbps.

\begin{figure}[ht]
 \begin{subfigure}[b]{0.5\linewidth}
  \centering
  \includegraphics[width=1\linewidth]{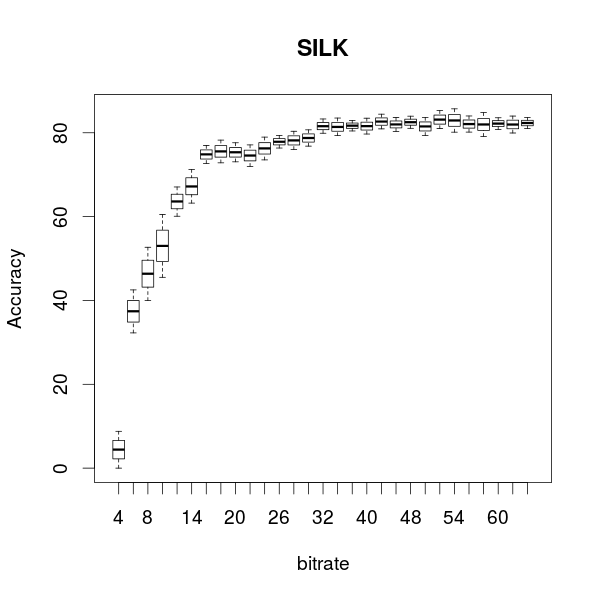}
  \label{fig:codec:silk}
  \vspace{-2em}
 \end{subfigure}
 \begin{subfigure}[b]{0.5\linewidth}
  \centering
  \includegraphics[width=1\linewidth]{codec-OPUS.png}
  \label{fig:codec:opus}
  \vspace{-2em}
 \end{subfigure}
 \begin{subfigure}[b]{0.5\linewidth}
  \centering
  \includegraphics[width=1\linewidth]{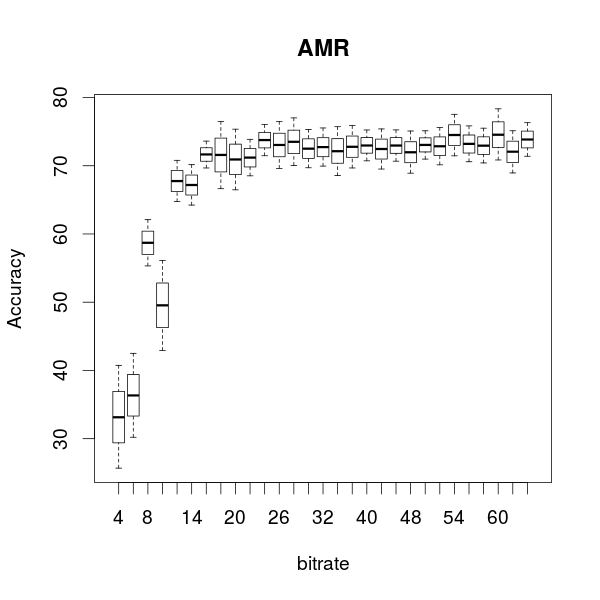}
  \label{fig:codec:amr}
  \vspace{-2em}
 \end{subfigure}
 \begin{subfigure}[b]{0.5\linewidth}
  \centering
  \includegraphics[width=1\linewidth]{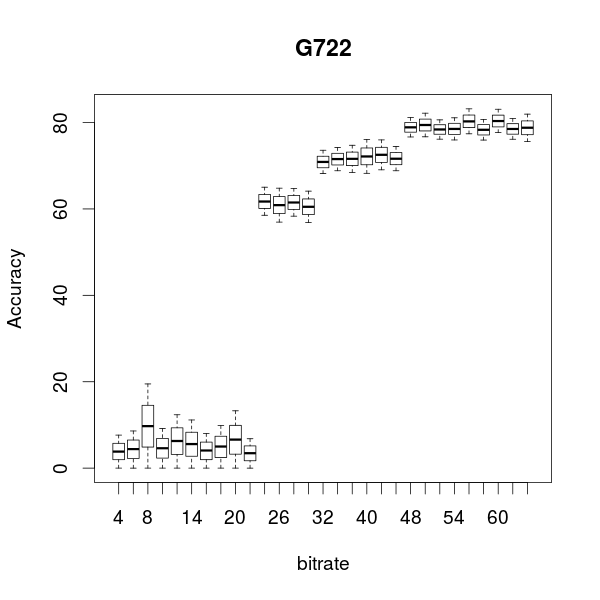}
  \label{fig:codec:722}
  \vspace{-2em}
 \end{subfigure}
  \caption{Location accuracy vs bitrate}
  \vspace{-1em}
 \label{fig:codec}
\end{figure}

Figure~\ref{fig:codec} shows the impact of bitrate on fingerprinting efficiency (detection rate) for various codecs. We observe a rather low efficiency ~38\%, at very low bitrates in the range of 6--10Kbps. At around 14--16Kbps, we observed an increase in attack efficiency to 73\%. This is interesting and linked to the increase in sampling rate to 12KHz around 10--12Kbps when SILK assumes a wider signal bandwidth of 6KHz. A steady improvement in attack efficiency is noted, as the sampling rate improves transmitting a greater part of the signal bandwidth to the receiver. Another increase (to 77\%) is noted at 24Kbps, when the codec switches to 24KHz sampling rate internally, also known as the superwideband mode in SILK parlance, stabilizing to 82\% at 40Kbps. At higher bitrates, SILK is able to support a wider band of frequencies, allowing a larger fraction of the signal features to be transmitted which improves fingerprinting.

\paragraphb{AMR-WB (Blackberry AIM): }
AMR-WB (Adaptive Multi-Rate Wideband) is an audio data compression scheme optimized for speech coding in telecommunications systems such as GSM and UMTS. It employs Linear Predictive compressive coding, just like SILK does in Narrowband mode. However, unlike SILK AMR employs LP throughout, to support signal frequencies from 50Hz to 7000Hz. AMR supports bitrates from 4.75 Kbps onwards up to 23.05 Kbps. AMR is somewhat dated as a codec in terms of design principles and applications, we use it as a baseline.

\paragraphb{722.1c:} is a low-delay generic audio codec~\cite{722.1c} standardized by the ITU. It is deployed widely in hardware-based VoIP devices particularly the Polycom series. It offers supports sampling rates above 32KHz offering a wider band than AMR, whilst supporting bitrates of 24, 32, and 48Kbps. Results in figure~\ref{fig:codec} show that a bitrate of at least 24--32Kbps is required to achieve reasonable fingerprinting success.

\paragraphb{Opus (Facebook Messenger/Zoom):} The Opus codec is a framework
for composing high quality codecs, namely SILK~\cite{silk} and
CELT~\cite{celt}. It operates in three modes: SILK mode, a new hybrid
mode, and CELT mode. In the SILK mode, it supports narrow to wide
frequency bandwidths, with relatively low-bit rates. The CELT mode is
a high-bitrate consuming codec offering a greater bandwidth than the
SILK mode. We observe a detection rate of 50\% at 10Kbps in
LP mode. At 12Kbps, we observe a significant improvement of 20\% in
fingerprinting efficiency to 70\% (which is in the realm of
usefulness). This is the threshold when Opus switches to Hybrid (wide
band) mode i.e from lossy to lossless compression, once again
confirming the importance of mid-range frequencies in the accuracy of
room fingerprinting. This is of interest, since it's meant to fill the
gap between LP mode and the MDCT mode. As the bitrate increases, the
signal bandwidth increases, leading to greater fidelity at the
receiver. At 14Kbps, the Opus codec shifts from LP to hybrid mode, entering a lossy compression stage onece again, resulting in reduced attack effectiveness
compared to the LP mode. At around 18Kbps, the codec recovers to
the same level as LP mode at 12Kbps. A second threshold increase is
noted at 20Kbps as the hybrid mode starts to support the
super-wideband frequency range. Gradual further improvement is noted
to 85\% which is fairly close to the baseline (no compression) figure
of 87\% accuracy. This is achieved when the bitrate is high enough
($>$ 48Kbps) to allow lossless compression in CELT mode
super-wideband.

A frame length of 20ms, at constant bit rate, was used in all
experiments. Opus supports short (2.5ms) and long (60ms) frame
lengths. The shorter the frame, the higher the bitrate. Further, Opus supports
redundant information, which improves quality at a cost of higher
bitrate allowing the decoder to recover against frame losses due to
random faults. In addition to frame length adjustment and redundant
information, Opus also supports multiple frame packetization. This
improves coding efficiency by reducing the number of packet headers
needed per second at the cost of additional delay. Overall, we have
focused our analysis on the impact of bitrate and assumed the network
path is free of significant variations in jitter and other error
conditions. We relax this assumption in Section~\ref{sec:jitter}.

\iftrue
 This is sustainable under the assumption that the
significant parameter is variable network bandwidth available to the
VoIP application resulting in variable bitrate. As part of future work
in the area, we plan comprehensive analysis involving other
parameters, namely redundancy, frame length, jitter, look ahead, and
training and testing on different conditions influenced by these
parameters.
\fi

\iftrue
\vspace{-1em}
\subsection{Impact of room occupancy}
Next we study the impact of time-variant properties of the location
such as the number of people present and the {\em speaker movement}.

Conti et al.~\cite{conti:2004:apl} showed that human bodies are good multi-directional reflectors assisting in the mixing of sound within a room. They demonstrated that the extent of reflection is solely described by the mass of the human which acts as a rigid water-filled ellipsoid. Absorption by the human body is largely dominated by the amount and type of clothing --- naked bodies will reflect entirely while thick winter coats will increase signal absorption. Since the fingerprinting attack is primarily sensitive to the absorption characteristics of the room, the expected impact is minimal in most circumstances. However, as reflectors, the scattering of sound from the human body can cause new influences. For instance, if occupants block a reflective surface such as a whiteboard, it could change the room's fingerprint.

To study the impact of room occupancy on location fingerprinting, we ran experiments on a set of 18 meeting rooms with various rates of occupancy -- this is the number of people in the room measured as the proportion of the maximum seating capacity of the room. We then tested VoIPLoc's efficiency at various rates of occupancy. For each room, we set up a VoIP call over Tor over a period of 8 to 12 minutes. We collected sound traces specifically for training by having each speaker read a standard script (with consent).

\begin{figure}[ht]
 \begin{subfigure}[b]{0.5\linewidth}
  \centering
  \includegraphics[width=1\linewidth]{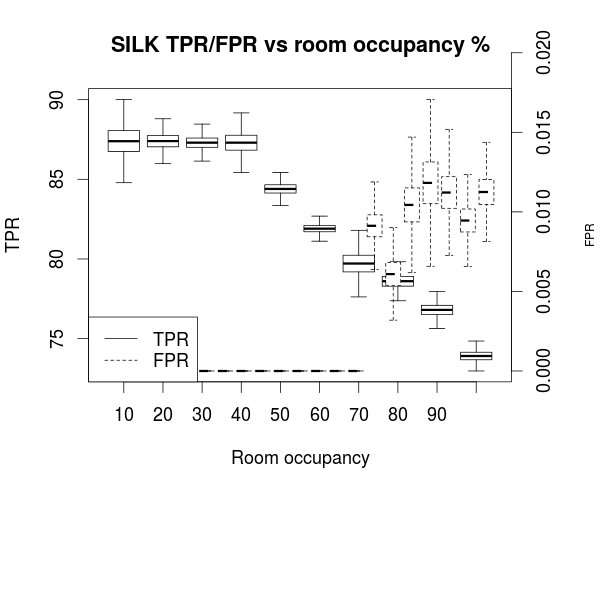}
    \vspace{-4em}
  \caption{SILK}
  \label{fig:occ:silk}
 \end{subfigure}
 \begin{subfigure}[b]{0.5\linewidth}
  \centering
  \includegraphics[width=1\linewidth]{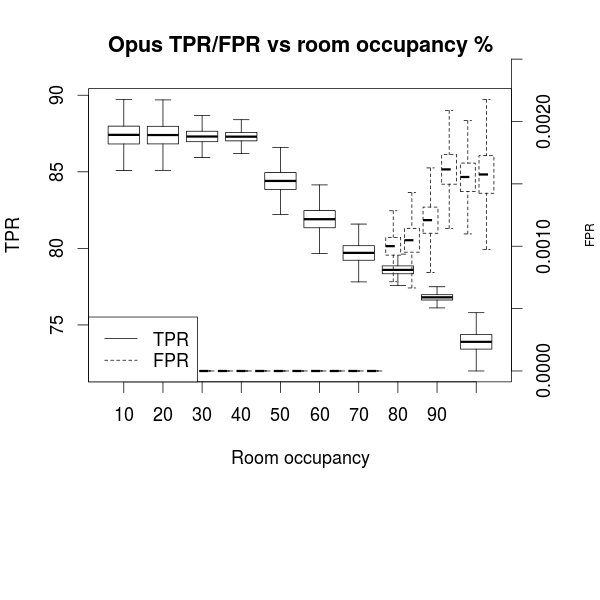}
    \vspace{-4em}
  \caption{Opus}
  \label{fig:occ:opus}
 \end{subfigure}
 \caption{Location accuracy vs room occupancy}
   \vspace{-1.5em}
 \label{fig:occ}
\end{figure}

Figure~\ref{fig:occ} shows that the detection rate gradually decreases at
around 50\% occupancy. We note a detection rate of over 85\% for all occupancies below 50\%, decreasing to above 70\% for 100\% occupancy. Our observations indicate that occupancy has impact but the attack is still a credible threat even at high occupancy rates. In the case of lecture and meeting rooms, all occupants were seated and the speaker was standing.

To confirm our hypothesis that attack accuracy is sensitive to absorption, we carried out another experiment. We examined the efficiency of attack when the VoIP sender (speaker) is surrounded by other people. For instance, when a hostage speaking whilst surrounded by kidnappers, or team participants huddled around a microphone during a conference call. We observed that the detection rate incurs a threshold decrease to 50\% after 40\% of the surrounding space is blocked. Further, as the surrounding space is progressively blocked, the impact of the walls on the room behaviour is {\em fully} replaced by the sound scattering properties of the occupants. In the worst case, only a minimal amount of reverberation is created. The detection rate does not decrease to zero because sound being a pressure wave can bend around obstacles although this attenuates the signal.

\vspace{-1em}
\subsection{Robustness to network jitter}
\label{sec:jitter}
VoIP traffic flows are routed over the Internet as a sequence of
packets. In the process, flows can experience variability in the
inter-arrival times of packets (jitter), experience loss of packets,
and variation in throughput due to dynamic router work-loads. Packets
that arrive too late at the destination are not played out (discarded).

Since packet delays and losses reduce audio quality, most codecs used by secure messaging systems implement a (packet) loss concealment strategy to maintain a perceptual level of voice quality despite any residual packet loss. Both OPUS and SILK codecs generate a replacement signal using the frequency spectrum of recent segments. For instance, substituting the missing signal with another signal with identical frequency spectrum, whilst replicating the pitch waveform from a recently received speech-segment signal. We note that jitter by itself does not affect the attack efficiency, since the packets arriving late can still be leveraged for fingerprint construction, although they are not played out. Hence, the focus of our analysis is on missing packets rather than delayed ones.

\begin{table}
 \centering
\scriptsize
\begin{tabular}{@{\extracolsep{1em}}lcccc@{}}\hline
{\bf Packet loss \%} & {\bf SILK (TPR\%/FPR\%)} & {\bf OPUS (TPR\%/FPR\%)}\\\hline
10 & 83.64/0.60 & 92.10/0.42\\
15 & 82.14/1.94 & 86.97/0.74\\
20 & 70.40/3.32 & 83.32/0.84\\
25 & 55.46/16.34 & 75.19/1.67\\
30 & 33.11/17.46 & 45.24/5.88\\
\end{tabular}
\caption{Impact of packet-loss on attack efficiency}
 \vspace{-4em}
\label{tab:jitter}
\normalsize
\end{table}

We introduced packet losses at various rates and observed changes in
attack efficiency using a configurable router. A Pica8 3920 SDN switch
was used to routing flows between source and destination pairs. The
switch was programmed to drop packets from the source-destination
flows at a selected rate of packet loss according to a poisson
distribution, which has shown to be a realistic assumption for standard network
traffic in applications such as arrival of HTTP/VoIP sessions~\cite{medhi2017network,muhizi2017analysis}.
For packet loss rates of 10\%, we find that the attack efficiency is
fairly high with low enough FPR and reasonable detection rates of
above 90\% in the case of Opus. In the case of SILK codec, the
detection rates are around 80\% for 10\% loss, which then reduce to
70\% for a loss rate of 20\%. For higher rates of packet loss, attack
efficiency is severely degraded in both cases to less than 50\%. More,
importantly we note that the FPR in Opus is relatively stable, being
less than 1\% until medium levels of loss (10\%), increasing only to
8\% for 30\% loss. The attack efficiency degrades faster when
operating via the SILK codec for increasing losses; beyond 10\%
loss-rates, FPR degrades to 14--17\% which is high. The reason for the
higher attacker efficiency via Opus is because of a dynamic jitter
buffer. When frames arrive after the length of the jitter buffer they
are discarded. In the case of Opus, the codec adapts to lossy network
conditions by embedding packet information into subsequent packets
allowing significantly better reconstruction rates and hence enhanced
attack efficiency in comparison with SILK.

To understand packet loss rates, Gu\'{e}guin et
al.~\cite{guguin:eurasip:2008} studied the mean opinion score, a well
respected metric to measure speech quality. Packet losses of less the
1\% are required to ensure no audible losses. A loss rate of 10\% is
considered poor with users being considerably annoyed at and beyond
this point. We note therefore that if the VoIP connection delivers
good to excellent voice quality then it is a fit candidate for
fingerprinting purposes with low FPR.

\iftrue
\vspace{-1em}
\subsection{Robustness to location diversity}
In previous sections, we have discussed the effectiveness of fingerprinting very similar locations which is the harder case for fingerprinting. However, the real world is much more diverse. Therefore we now add {\em diverse} locations with very different characteristics to the fingerprint database and evaluate the effectiveness of the technique with this change. We selected locations with markedly different acoustic characteristics owing to significant differences in physical size, shape, volume, and construction material. Our goal is to understand whether sound samples from the same room can be linked given a population of VoIP traces from a diverse range of indoor venues. Our dataset consists of sound samples collected in locations given in Table~\ref{tab:roomdb}.

Location diversity also gives us the opportunity to study additional parameters. Reverberation length (the length of the reverberant component) is shorter in meeting rooms as opposed to warehouses or lecture rooms. Diverse room sizes also induce variability in reverberation length so it's important to accurately estimate this. Additionally, we also consider the impact of utterance length i.e the amount of time for which signal power is above the threshold of silence. An utterance corresponds to one or more words spoken together such that the signal amplitude does not fall to the noise floor (0 dB). Long utterances have higher aggregate signal power compared to shorter utterances which makes fingerprinting easier.


\begin{table}
 \centering
 \scriptsize
\begin{tabular}{|p{2cm}|p{5cm}|}\hline
Room         & Description \\\hline
Building Atrium (enclosed courtyard) & Semi-enclosed Atrium of the CS department. The courtyard is enclosed on three sides by partition board walls and on the third side by a glass wall. Volume is roughly $24000m^3$.\\
University Sports Centre & A large sports hall in the sports center of the university. The reverberation length is fairly long. Volume is 12000$m^3$.\\

Typing Room & Partitioned typing room in the British Museum.\\

Office rooms & 79 rooms from a university CS department\\

Stairway & Stairway within an office building.\\

Concrete studio & Bare room with plastered walls, concrete floor, and concrete ceiling.\\

Underground car park & Cemented, with pillars, 8500$m^3$ \\

Domestic living room & Wood and lathe, 1100$m^3$.\\

Wood paneled room & Wood paneled studio room, 1400$m^3$\\

Stanbrook Abbey Malvern & Large hall within a former monastery, 61000$m^3$.\\

Factory warehouse & Large empty warehouse inside the Cadbury factory, 4800$m^3$.\\\hline
\end{tabular}
\caption{Indoor locations considered}
\normalsize
 \vspace{-3em}
\label{tab:roomdb}
\end{table}


\begin{figure}[ht]
 \begin{subfigure}[b]{0.5\linewidth}
  \centering
  \includegraphics[width=1\linewidth]{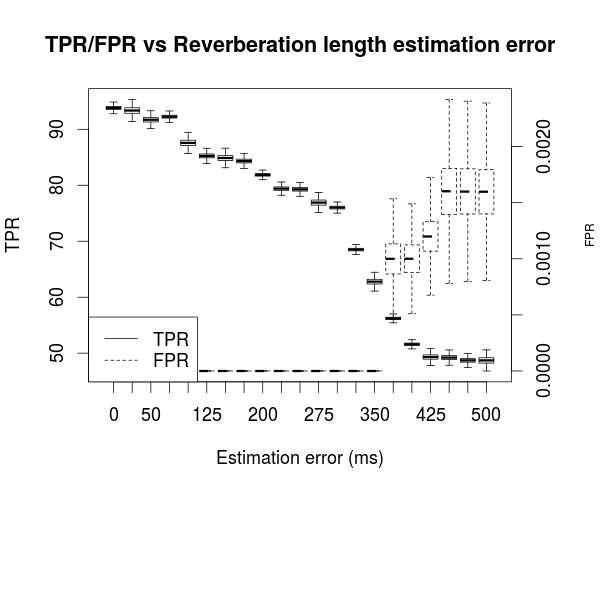}
    \vspace{-4em}
  \caption{Impact of RT estimation error}
  \label{fig:rterror}
 \end{subfigure}
 \begin{subfigure}[b]{0.5\linewidth}
  \centering
  \includegraphics[width=1\linewidth]{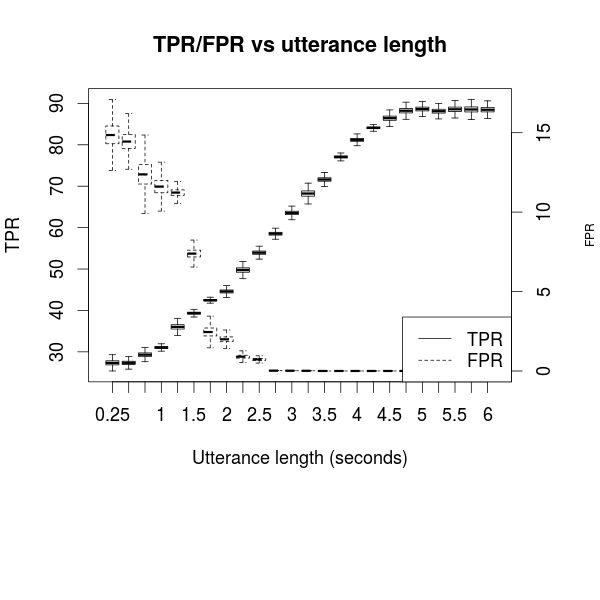}
    \vspace{-4em}
  \caption{Impact of utterance length}
  \label{fig:utterance}
 \end{subfigure}
 \caption{Impact of Parameters on Accuracy (Opus)}
   \vspace{-1.5em}
 \label{fig:impact:opus}
\end{figure}

\paragraphb{Parameter estimation} The classifier depends on accurate estimation of reverberation time to isolate the reverberant component. In practice, this parameter must be estimated without having any information other than the signal itself. Our method estimates this parameter, with some errors. The amount of error depends on the noise level within the signal. Hence, we evaluate how poor reverberation-time estimation impacts overall accuracy. We compare the error in room identification as a function of estimation error. We vary the error in milliseconds ($ms$) from the ground truth. The results are shown in Figure~\ref{fig:rterror}. The results show that if there is a very small error ($100ms$) in estimation, then location identification accuracy is above 87\%. If the estimation error is larger ($200ms$---$300ms$), then the accuracy drops to around 75\%. In indoor offices and residence halls, we found the estimation error to be within $60ms$. In the stairway, the error was $50ms$, and around $400ms$ in the warehouse. The key insight here, is that the quality of separation between datapoints corresponding to different classes (separating hyperplane) was of sufficiently high quality as to enable a high detection rate that decayed linearly with quality of capture of the reverberant component (higher the RT estimation error, lower the quality of input datapoint into the classifier, hence lower the classification output). A second insight is that we observe a threshold effect in the false-positive rate; the false-positive rate is zero until RT error is $350ms$ but experiences a threshold increase at $375ms$. Until the threshold value, quality separation between the various data-point categories (location classes) cushions the impact of the estimation error on the false-positive rate. Beyond the threshold, the input fingerprint candidate vectors are simply noise and get categorized together leading to a significant rise in the false-positive rate.

\paragraphb{Utterance length}. A second parameter is the minimum signal power conveyed by the speaker in a single word or sentence (utterance). If the utterance length is very small, the signal strength is too low for fingerprinting. A human utterance is a consecutive set of speech segments. During data collection, we observed that the length of an utterance varies in duration (a well established fact in the literature). Our classification technique works by extracting the statistics of the reverberant component across the length of an utterance. Conversations with long utterances increase accuracy of the classification. To understand the impact of utterance length, we used the conversations with varying lengths and observed classifier accuracy. 
In figure~\ref{fig:utterance}, we observe that if the length of utterances is greater than 3.8 seconds, we obtain over 78\% accuracy for the classifier, while for utterances lasting less than $1$ second (monosyllable words) the accuracy is less than 28--30\%.
\fi

\iftrue
\vspace{-1em}
\subsection{Scalability}
The ability to fingerprint a location is only as useful as the number
of locations that can be uniquely fingerprinted. One concern is that
VoIPLoc may perform less accurately with a larger number of rooms as it
may become easier for fingerprints to ``collide''.

\if JOURNAL
Allen and Berkeley~\cite{allen:jacoustsoc:1979} famously proposed the
Image-Source Method (ISM) to generate an acoustic recording for a
virtual room. A virtual room is constructed with given geometric
dimensions and reflection coefficients for various reflective surfaces
within the room. The {\em characteristic function} of the room defines
how sound energy is reflected across the considered room for up to
$n$-recursive reflections. Once the characteristic function is
available, audio traces within the room are generated by convolving
the characteristic function with (direct-sound) speech. The speech
component used here must be free of acoustic reflections i.e the
direct sound component. The location-characteristic function is also
known as the Room-Impulse Response function in signal processing
literature. Allen and Berkeley were the first to propose a full
implementation of their image-source method. However, at the core of
the ISM lies a histogram that registers acoustic impulses versus time
generated at a given recording location within the room, given all
reflective sources. The use of a histogram to capture acoustic energy
leads to rounding of signal power values to the nearest bin centre
value, resulting in a build up of error --- a synthetic reverberant
signal would differ from its real-world counterpart (were the
synthetic rooom to be constructed and measured in the real
world). Thus such a synthetic model would be unserviceable for our
scalability study. To overcome these difficulties, an alternative
`high-fidelity' technique by Lehmann and
Johansson~\cite{lehmann:jacoustsoc:2008} uses frequency-domain
simulation of sound reflections, which allows relatively accurate
representation of delays that are not integer multiples of the
sampling period. To choose an appropriate method for generating our
synthetic dataset, we compared the reverberant signal generated via
the synthetic method with the actual reverberant component from the
ground-truth in our first dataset~\ref{sec:eval}. We compared the
aggregate (normalised) signal power across all frequencies for
synthetic versus actual signals for the same room, for 79 different
rooms. Figure~\ref{fig:synerror} plots the aggregate (normalised)
signal power across all frequencies for synthetic versus actual
signals for the same room. With some deviation, we can see that the
expected vs actual signal values are closely correlated, which means
that the simulation method we've chosen is a reliable approximation of
fingerprinting the corresponding real-world location. The results are
for 79 different rooms. We can found that the Lehmann-Johansson method
provides the closest approximation of the actual signal, which forms
the basis for our choice of the method to create a synthetic dataset.

\begin{figure}[!ht]
  \centering
  \includegraphics[width=1\linewidth]{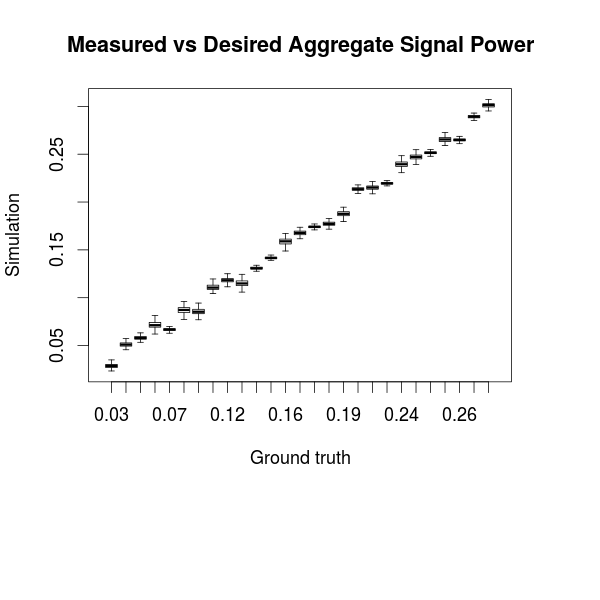}
  \caption{Plot of ground truth vs measured signal power}
  \label{fig:synerror}
 \end{figure}

\fi

To get a sense of performance over a larger number of locations we
used a large-scale synthetic location dataset comprising rooms with
realistic room and furniture layouts. Each room is manually filtered
for its realism by a set of three volunteers recruited via Mechanical
Turk. If at least two volunteers vote in favour of a synthetic room
layout, it is considered realistic. After filtering, the number of
realistic furnished rooms in the dataset are $404$,$058$. We generated
the synthetic traces as follows: for each room, we generated the
room's characteristic function using the Lehmann-Johansson
method~\cite{lehmann:jacoustsoc:2008}. We convolved the function with
an anechoic audio recording to generate the audio traces containing
direct sound, overlapped with reflections, further overlapped with a
reverberant component. Each audio trace is therefore a realistic
emulation of the room constructed with real building materials with
the exclusion of background noise.  The Lehmann-Johansson method uses
the geometric dimensions (length, breadth, and height) of a room,
along with speaker position and orientation, and the reflection
coefficients of the room surfaces. We set the reflection coefficients
using industry standard values for various labelled
materials~\cite{vorlander:2007} as per the dataset. As with the
real-world dataset in Section~\ref{sec:rdata}, we vary orientation and
position. We used the voices of 16 speakers from the NXT
corpus~\cite{nxtcorpus} (also used in section~\ref{sec:rdata}, to
obtain $1.4$ billion. Each recording is transmitted via a secure
messaging client (Skype) communicating to a recipient over a Tor
circuit. 
The location fingerprint is computed over the audio data at the
recipient.  The traces are then input to the classifier. As before
(\ref{sec:rdata}), we partition the dataset into $k$ non-overlapping
subsets, with one subset being used for training and $k-1$ for
testing. Each subset contains 5 traces per location. The results are
averaged over $k$ iterations as in a $k$-foldover cross-validation.

\begin{table}[!htb]
	\begin{minipage}{.48\linewidth}
		\centering
 		\scriptsize
		\begin{tabular}{@{\extracolsep{0em}}lccc@{}}\hline
 			Codec & Room count & \% FP & \% Detected \\\hline
 			SILK &  100 & 0.000  &  87.57\\
    			&  1000 & 0.047  &  87.65\\
    			& 10000 & 0.042  &  88.76\\
    			& 100000 & 0.008 &  88.93\\
    			& 404058 & 0.007 &  89.98\\

			Opus  &  100 &  0.001  &  90.76\\
    			& 1000 &  0.037  &  91.28\\
    			& 10000 & 0.024  &  91.41\\
    			& 100000 & 0.003 &  91.52\\
    			& 404058 & 0.003 &  91.77\\ \hline

		\end{tabular}
		\caption{Scalability results on 404,000 rooms}
		\normalsize
		\label{tab:scale1}
	\end{minipage}
	\begin{minipage}{.48\linewidth}
		\centering
		\scriptsize
		\begin{tabular}{@{\extracolsep{0em}}lccc@{}}\hline
			Codec & Room count & \% FP & \% Detected \\\hline
			SILK   &  1000 & 0.053 &  83.24\\
     		& 10000 & 0.014 &  84.38\\
     		& 100000 & 0.009 &  84.59\\

			Opus   &  1000 & 0.026 &  88.27\\
     		& 10000 & 0.018 &  89.36\\
     		& 100000 & 0.007 & 89.39\\\hline
		\end{tabular}
		\caption{Scalability --- results with partial traces}
		\normalsize
		\vspace{-3em}
		\label{tab:scale2}
	\end{minipage}
	\vspace{-2em}
\end{table}

\fi

\iftrue
Overall, we found that VoIPLoc scales well with the size of the
location-database, with performance remaining stable as the number of
locations increases. For example (in Opus), the false-positive rate (FPR)
for 10000 rooms is $0.02\%$ (Table~\ref{tab:scale1}), while for 100000
rooms this rate is $0.003\%$. The FPR
decreases roughly by a factor of $10$, which is equal to the scale-up
factor between the two experiments, indicating that the actual number
of false-positives remains the same.

\paragraphb{Training only with partial traces:} In the experiments we
have performed so far, the attacker has access to audio traces
generated from various victim positions and orientations. However, in practice such a
diverse coverage of a room's acoustic characteristics may not be
available for sampling. It is therefore useful to evaluate how well
VoIPLoc would work with datasets where only a fraction of the possible
audio traces are known for training. The training set is as follows:
conservatively, we assumed the victim is at one randomly chosen
position of nine and just three randomly chosen orientations out of
twenty. This gives reasonable room for head movment in a VoIP
session. We therefore removed 97\% of the audio traces from our
dataset. Table~\ref{tab:scale2} documents the effects of partial
data. While the detection rate falls, VoIPLoc still compromises the
location-privacy of over 83--88\% of users with high reliability. In
practice, the number of audio traces available per location will
likely be higher.

\begin{table}
\centering
\scriptsize
\begin{tabular}{@{\extracolsep{1em}}lccc@{}}\hline
Codec & Missing location count & \% FP & \% Detected \\\hline
SILK   &  101000 & 0.049 &  78.46\\
     &  202000 & 0.040 &  79.53\\
     &  303000 & 0.040  & 79.29\\
     &  383800 & 0.047 &  79.80\\

Opus   &  101000 & 0.019 &  83.76\\
     &  202000 & 0.024 &  83.03\\
     &  303000 & 0.026 &  83.95\\
     &  383000 & 0.026  & 83.81\\\hline

\end{tabular}
\caption{Open-world -- results with 404k locations with both unlabelled and partial traces}
\normalsize
\vspace{-3em}
\label{tab:openworld}
\end{table}

\paragraphb{Effects of unlabelled data:} Up until now, we have trained
VoIPLoc with a label for each location it can expect to be tested
with. However, in the real-world this won't be the case since all
possible locations won't have been fingerprinted. It is therefore
important to consider how missing locations will impact VoIPLoc's
performance. We simulate testing in the open-world setting by removing
a fraction of locations from the training phase whilst retaining them
during the testing phase. Table~\ref{tab:openworld} documents the
impact on VoIPLoc's performance. We removed an increasing number of
location labels from fingerprints in the training set but retained
them as unlabelled fingerprints in the testing set. Broadly, we find
that algorithmic efficiency is not impacted by low to medium levels of
noise from unlabelled locations. Notably the FPR is
26--50 per 100000 locations. Efficiency reduces by around 10\%
compared to the fully labelled set. We posit the unlabelled set is
contaminated with fingerprints that are actually positive, causing a
reduction in the efficiency of the classifier. To reduce the impact of this issue, the pre-filtering step in Section~\ref{sec:classification}
removed positive fingerprints from the unlabelled set which prevents
large-scale contamination.
\fi



\vspace{-1em}
\section{Discussion}
\label{sec:discussion}

Our attack technique allows participants of a VoIP call to extract
location-reflection characteristics from recorded audio to generate a
fingerprint.

  Our experiments confirm the hypothesis that the reverberant sound
 component can be used to generate location fingerprints, with high
 reliability and low false-positive rates of detection. We evaluated with rooms of identical geometry which are
 differentiated only by the customisation introduced by their
 occupants, such as the placement of monitors and the number of books
 on their shelves. The attack technique uses a deep NMF-SVM
 classifier, which was trained on a few samples per room, and then
 tested extensively against samples recorded in different parts of
 the location. This indicates that a fingerprinting technique can be
 used to reliably link an audio traces recorded at different parts of
 the same location. Given the low false-positive rates ($0.003\%$),
 we documented the location accuracy in terms of the detection rate
 alone. While we expected aggressive audio compression employed by
 the codecs to significantly damage the detection rates, we found
 that low-bitrate codecs such as SILK and Opus carry out an important
 function that improves detection rate: they remove background noise
 that negatively influences detection. In most cases, the steady
 state detection rates are between 60\% and 88\%, with a room
 occupancy (\% of maximum seating capacity) of less than 50\%, and a
 reliable network connection with 5--10\% network jitter.

 VoIPLoc does not depend on background sounds within a location or the
 voice of a specific speaker. Thus passive countermeasures such as
 filtering techniques will have little impact on attack efficiency
 since the fingerprint is computed over a basic VoIP-channel property
 --- delivering the speaker's voice to the receiver with integrity.

\iftrue
\paragraphb{Degrees of freedom:} Scalability of the fingerprinting process is
an important aspect to study. We evaluated VoIPLoc against a synthetic
dataset comprising $404$,$058$ rooms. Our experiments show that
VoIPLoc can scale to a large number of locations with a couple of audio samples per room. Thus we can say that VoIPLoc presents at
least five degrees of freedom to the attacker.
collect at scale.

\paragraphb{Tradeoffs between compression and privacy:} The room
structure and its interaction with compressive techniques employed by
modern codecs plays an important role in the communication of
characteristic information that can be leveraged for
fingerprinting. At the same time, the benefits of compression such as
reduced delay penalties are an important incentive for their use in
VoIP design. As part of future work, we plan to study the tradeoff
between compression and the ability to avoid location-detection, and
whether there exist fundamentally stealthy codecs that can mask
fingerprintable information but that are also serviceably compressive.
\fi

\paragraphb{Location confirmation:} VoIPLoc may be used in conjunction
with macro-geolocation techniques such as
PinDr0p~\cite{bala:ccs:2010}, as a way to combine coarse-grained and
fine-grained tracking.

\paragraphb{Countermeasures:} Defenders may consider a number of
approaches. First, defenders may use acoustic jitter to damage
fingerprint information. For instance, a constant amplitude signal a
the room's characteristic frequencies between 50Hz and 2KHz (the
discriminating subset of the location fingerprint) can cause a
significant decrease in VoIPLoc's performance. This is essentially an
acoustic jamming strategy which will deny access to the reverberant
component of the channel to the attacker (receiver). On the other
hand, any acoustic interference strategy will need to avoid jamming
the communication channel itself or causing substantial
disruption. However this is hard to achieve as even small amounts of
audible noise will negatively impact voice quality (hence unlikely to
be deployed). Alternately, network jitter can be used to induce packet
latencies encouraging standard codec implementations to drop packets
containing reverberant components. If accurately executed, this
countermeasure could be fairly effective in preventing the sender from
extracting a credible room fingerprint. As much as it would be
effective against a standard implementation, the attacker could retain
late packets (instead of dropping them) and access the reverberant
component. Further, the reverberant component may be encoded into
other packets as standard codec implementations often encode previous
audio data into transmitted packets to mitigate packet losses.

\iftrue

\paragraphb{Classifier design:} VoIPLoc uses a deep NMF-SVM classifier. It
combines techniques from DNNs (Deep Neural Networks) with a robust
classical approach for finding decision boundaries. From DNN
literature, we used multiple-layers, pooling, and normalisation, which
are among the promising components of deep neural
networks. Multiple-layers enable fine-grained partitioning between the
first, second, and multi-order reflections via hierarchical
decomposition that involves no training of weights (hence no
backpropagation); Pooling reduces the impact of noisy audio traces;
and, normalisation enables comparison across audio traces by
normalising out the effect of varying amplitudes of direct sound. The
VoIPLoc classifer architecture has multiple decomposition layers with
the final layer composed of support vectors. The reason for using SVM
as the final layer is that VoIPLoc requires a robust classifier that
works with small datasets. Unlike SVMs, DNNs require large training
datasets~\cite{du2018samples} which are not available in our problem
setting ---- one may need to distinguish just ten locations from a
large number of unlabeled location fingerprints. To increase the
quantity of training data, data augmentation techniques are commonly
used for DNNs. However, each network of neurons will still require a
significant amount (eg. ~1000 samples per class for Imagenet
classification~\cite{eccv:2016:imagenet}) of labeled data for it to
train before data augmentation can assist. Applying DNNs is therefore
challenging in the call provenance problem, where only a few tens of
samples per location are available for training in the best case and a
handful in the typical case.
Aside from poor performance over small datasets, a second reason for
choosing SVM is the appropriateness of the tool. DNNs are appropriate
for high-dimensionality problems such as image
classification~\cite{eccv:2016:imagenet} where the feature set for a
120$\times$120 pixel RGB image is 43200, and feature selection is left to the
classifier. This isn't the case with VoIPLoc, where the attack is
based on a specific feature of the audio traces, namely the
reverberant component for location inference. Thus acoustic location
fingerprinting based on the reverberant component would not draw
on one of the main strengths of DNNs -- the ability to perform better
in a high-dimensionality setting as compared with SVMs~\cite{nature:2015:rl}.

\paragraphb{Bagging vs Boosting:} An alternate approach to combine the
output of different classifiers is {\em Boosting} (classifiers with
votes weighted by accuracy) as opposed to the {\em Bagging} approach
used by VoIPLoc (weak classifiers trained on a subset of the training
data).  Boosting requires a large training set which is typically
unavailable in room fingerprinting whereas Bagging can be successfully
trained with a small sample set.  Also, Bagging performs relatively
better with noisy data than Boosting
approaches~\cite{kotsiantis:air:2011}.  At the same time, Bagging
approaches increase complexity. So are they worth the additional
complexity? We tested VoIPLoc with and without Bagging.  We observed
that across the experiments, there was a 5\% improvement in TP
detection rate and a small decrease in the FPR. This indicates that
while the noise tolerance of Bagging contributes in a positive manner,
the improvement achieved is not very significant ($<10\%$), and
therefore we thing it is not worth the additional complexity.

\paragraphb{Open-world vs closed-world:} We evaluated VoIPLoc in both
a closed-world environment as well as an open-world scenario
(unlabeled fingerprints). As long as the proportion of unknown
locations is less than 85\% of the fingerprint database, the
FPR is serviceable. Beyond that, unlabelled data
penetrates through the contamination-resistance filters, decreasing
detection efficiency.

\fi

\paragraphb{Indoor vs. Outdoor application:} Outdoor locations typically demonstrate poor reverberance characteristics, and require excitation signals of higher amplitude than human voices whilst in a VoIP conversation. For this reason, the applicability of this work is primarily of significance in fingerprinting indoor locations.

\vspace{-1em}
\section{Related work}
\label{sec:relwk}

A number of works have attempted to derive location information via
side-channels. The techniques can be broadly classified into passive
approaches and active approaches. VoIPLoc is the first
passive technique to achieve fingerprinting via echo-location
characteristics of call origin.
None of them are designed to work
specifically against anonymous VoIP channels.

\vspace{-1em}
\subsection{Passive approaches}
A number of passive approaches use static features of background
noise. SurroundSense~\cite{azizyan:2009:mobicom} combine sound
amplitude with camera and accelerometer inputs to distinguish between
indoor locations via overall ambiance (sound, light, and
decor). ABS~\cite{tarzia:2011:mobisys}, identifies a location using
low-frequency background sounds (computers, fans, buzz of electrical
equipment). Next, we look at passive approaches that use dynamic
features of background noise.  Kraetzer et
al.~\cite{kraetzer:2007:wms} propose a location identification
technique that relies on repetitive patterns of music played in a
location.  Lu et al.  (SoundSense)~\cite{lu:2009:mobisys} generalises
this to use background sounds such as passing trains and associates
each location with a set of identifiable background sounds.  Usher et
al.~\cite{usher:2007:ieeetaslp} generalized this a bit further by
replacing music with the voice of a single human speaker.  Malik et
al.  carried out a small study over four very differently sized rooms
and showed that differently sized rooms had different length and decay
rate of reverberation~\cite{malik:2010:icassp}. Parhizkar and others
extend this to echo-based approaches that leverage background sound as
impulse
signals~\cite{parhizkar2014single,eronen:2006:ieeetaslp,lane2015deepear,davis:1980:ieeetaslp,chu:2009:ieeetaslp,kotropoulos2014mobile}. Vaidya
et al.~\cite{vaidya2019whisper} describe re-identification attacks,
wherein an attacker can infer location data from underlying audio, by
analysing packet size distribution.  While these approaches can
distinguish a street from an airport it isn't serviceable for
confirming the location of a VoIP user. The main challenge, as before,
is that aggressive compressive encoding filters out all background
signals.


\if false
In the context of multimedia copyright techniques, audio
fingerprinting (also called acoustic fingerprinting) is used to map an
audio recording to a unique identifier. These
techniques~\cite{zmudzinski:2008:wms,malik:2010:icassp,george2015scalable,son2020robust} 
are suitable for searching for a noisy song-snippet within a music
database. These techniques have little to do with location
fingerprinting, and instead focus on fingerprinting audio content
(identifying the speaker or the music clip). Interestingly however, the authors
in \cite{zakariah2018digital} describe that recordings from different locations can be used to
establish the stability of the electronic network frequency (ENF) used in
forensic analysis for digital multimedia -- which is arguably the most reliable
method for integrity verification and forgery localisation.
\fi

Finally, within passive approaches, we review a number of
complimentary techniques that leverage network characteristics instead
of acoustic side-channels for localisation. Wang
et al.~\cite{wang2016csi} fingerprint locations based on channel-state
information (CSI) from WiFi network interface cards. The use of CSI
for localisation shows to be more successful than previous machine
learning approaches which typically use stored received signal
strengths (RSS) that have been described to have high variability for
fixed locations and are very
coarse~\cite{youssef2005horus,podevijn2018performance}.
There are also works in the telephony-spam and mobile-fraud detection
literature on mobile device
provenance~\cite{kotropoulos2014mobile,mehdi2016optimising} which is
complementary to VoIPLoc's location provenance.
PinDrop~\cite{bala:ccs:2010} leverages distinguishing acoustic
characteristics arising from the device used, such as peak activity,
choice of codecs, and double talk. It also uses characteristics
induced by the network path on the codecs involved such as packet-loss
rates in different networks. Similar ideas of leveraging packet-delay
metadata has been proposed by Abdou et
al.~\cite{abdou2018internet}. Unfortunately, Pindrop's fingerprint
(and Abdou's work) being based on path characteristics, is vulnerable
to the effects of network components such as Tor routers. VoIPLoc is
less impacted since it doesn't fingerprint using network
qcharacteristics as it directly focuses on location
characteristics. While empirical testing with Tor has been carried out
in this work, VoIPLoc's ability to remain undeterred where other
solutions such as OnionPhone~\cite{gegelonionphone} ({\em successor}
of TORFone~\cite{torfone}) and Phonion~\cite{heuser2017phonion} may be
employed has to be experimentally verified. Further, PinDrop's
fingerprint is coarse grained by its very nature of being path and
equipment centric characteristics. VoIPLoc's fingerprint is
fine-grained being a function of the room's physical
characteristics. This suggests that a unified fingerprinting approach
that combines network-path, equipment, and room echolocation
characteristics may be the best way forward.
Malik et al.~\cite{malik2012recording} was the first to apply machine
learning to the problem (using SVM) but their fingerprinting technique
suffered fundamentally. It measured the power spectrum (MFCC
coefficients) without separating out reflections leading to two
problems. First, contamination by background noise. Second, the
fingerprint becomes a function of a specific spot within a room owing
to contamination by direct-sound and early-reflections.




\vspace{-1em}
\subsection{Active approaches}
\label{app:active}
Recent works on privacy and acoustic channels include sensory malware,
notably Soundcomber~\cite{schlegel:ndss:2011} and its variants which
extract sensitive information such as credit card numbers from
acoustic channels on a smartphone. There have also been reports of
malware that communicate across air gaps via ultrasound
squeaks
~\cite{hanspach:jcm:2013}. A research paper based on this idea is
SonarSnoop~\cite{cheng2018sonarsnoop} which claims undetectability
 as no noise or vibrations are
induced. However, health and well-being apps~\cite{Zoosh,usoundapp}
that record ambient 'energy' levels on a per-source basis will
register these ultrasound beeps.
A number of other works leverage the reverberation component for a
function similar to fingerprinting. The state-of-the-art technique in
this space for signal separation --- isolating a high-resolution
version of the reverberant component~\cite{dokmanic:2013:pnas} --
requires four microphones spaced exactly one meter apart. A
first-order approximation using two microphones can be obtained using
the technique of Pradhan et al.~\cite{pradhan2018smartphone}. VoipLoc
is the first to achieve credible separation of signal components to
isolate the reverberant component using a single microphone without
the need for a synthetic impluse signal.
In the audible range, Shumailov et al.~\cite{shumailov2019hearing}
propose an active acoustic side-channel attack using the device
microphone to record sound waves which present over a touch-screen
when it has been tapped by a user. A more sophisticated approach is to
inject an optimal (short-high amplitude) impulse signal into the
target location and measure the impulse
response~\cite{naylor:1993:aa,tsingos:2004,tarzia:2009:ubicomp} apply
forensic measurement techniques to develop high-fidelity acoustic
model of a location to simulate the effects of a room over a given
anechoic signal~\cite{farina:01:aes}. Murgai et
al.~\cite{murgai2017blind} estimate the decay time and amplitude of
the reverberant component to estimate room volume but not necessarily
a specific room. In comparison, VoIPloc can differentiate between
rooms of identical geometry by isolating the frequency components
within a reverberant component of audio recorded with a single
microphone.

\vspace{-1em}
\section{Conclusion}
Applications supporting and accepting voice based communications are
very popular. Humans record and exchange audio data on a planetary
scale. VoIP is a popular and important application used by dissidents,
police, journalists, government, industry, academics, and members of
the public. Thus privacy for VoIP applications is an important
requirement. Beyond VoIP, voice-control is an increasingly popular
method of user-device interaction in smart devices, which might reveal
the fine-grained information about a user's location down to which
part of the building they occupy.

Location information is embedded into human voice due to acoustic wave
propagation behaviour, which forms the basis of location
fingerprinting --- the reflections of direct-sound interfere with each
other and with direct-sound resulting in a rich interference pattern
carried by encoded human voice. Given the wide usage of smartphones
and VoIP tools, among the wider public to record and transmit audio,
this work has important implications for anonymous VoIP communication,
and more generally on user expectations of privacy within their homes
as fine-grained user-location information can be derived from
audio-interfaces to IoT devices. In terms of machine learning, while
DNNs are popular, we learned that they are not the best option universally.
In the call provenance challenge, given the sparse availability of traces
to compute a fingerprint, less is more. Consequently, a hybid SVM-centric
approach does better than a neural network approach.

\begin{acks}
The authors are indebted to the numerous volunteers who helped with
the field experiments from the University of Birmingham CS department
and School of Informatics, University of Edinburgh. A particular note
of thanks to Jon Weekes. The authors are thankful to the reviewers
for their insightful comments.  The first author is
supported by a UKIERI grant (UKIERI2018-19-005). The second author is
funded by EPSRC (EP/11288S170484-102) and NPL's Data Science program.
\end{acks}

\bibliography{paper,p3ca} 
\bibliographystyle{plain}

\end{document}